\documentstyle[12pt]{article}
\begin{document}
\begin{center}{\large{\bf Conformal Form of Pseudo-Riemannian Metrics by Normal Coordinate Transformations}}
\end{center}
\vspace*{1.5cm}
\begin{center}
A. C. V. V. de Siqueira
$^{*}$ \\
Departamento de Educa\c{c}\~ao\\
Universidade Federal Rural de Pernambuco \\
52.171-900, Recife, PE, Brazil.\\
\end{center}
\vspace*{1.5cm}
\begin{center}{\bf Abstract}

In this paper we extend the Cartan's approach of Riemannian normal
coordinates and show that all n-dimensional pseudo-Riemannian
metrics are conformal to a flat manifold, when, in normal
coordinates, they are well-behaved in the origin and in its
neighborhood. We show that for this condition  all n-dimensional
pseudo-Riemannian metrics can be embedded in a hyper-cone of an
n+2-dimensional flat manifold.  Based on the above conditions we
show that each n-dimensional pseudo-Riemannian manifold is conformal
to a n-dimensional manifold of constant curvature. As a consequence
of geometry, without postulates, we obtain the classical and the
quantum angular momenta of a particle.
  \end{center}

 \vspace{3cm}

${}^*$ E-mail: acvvs@ded.ufrpe.br

\newpage

\section{Introduction}
$         $

Conformal spaces are very important in geometry and physics.
Researchers pay special attention  to them and there are several
important results based on or associated with conformal geometry
\cite{1}, \cite{2}. In this paper we present, in detail, results of
Cartan, \cite{3}, \cite{4}, \cite{5}, and make a simple extension
that implies a new consequence: all n-dimensional pseudo-Riemannian
metrics are conformal to a flat manifold, when, in normal
coordinates, they are well-behaved in the origin and in its
neighborhood.
\newline
This paper is organized as follows. In Sec.$2$ we present normal
coordinates and elements of differential geometry. In Sec.$3$  we
continue the geometric approach. In Sec.$4$ we show that all
well-behaved n-dimensional pseudo-Riemannian metrics in origin and
in its neighborhood, in normal coordinates, are conformal to a
n-dimensional flat manifold and to a n-dimensional manifold of
constant curvature. This result is used in the Cartan's solution for
a space of constant curvature. In Sec.$5$ we present more
differential geometry by introducing normal tensors to build the
Cartan's solution for a general pseudo-Riemannian metric. In Sec.$6$
we make an embedding of all n-dimensional pseudo-Riemannian metrics
that obey previously presented conditions into a hyper-cone of a
flat n+2-dimensional space. In Sec.$7$, we make an embedding of all
n-dimensional pseudo-Riemannian manifold of constant curvature in a
n+1-dimensional flat manifold, obtaining, without postulates, the
quantum angular momentum operator of a particle as  a consequence of
geometry.
\renewcommand{\theequation}{\thesection.\arabic{equation}}
\section{\bf  Normal Coordinates}
$         $
 \setcounter{equation}{0}
 $         $
 \setcounter{equation}{0}
 $         $

In this section we briefly present normal coordinates and  review
some elements of differential geometry for an n-dimensional
pseudo-Riemannian  manifold, \cite{3}, \cite{4}, \cite{5}.
\newline
Let us consider the line element
\begin{equation}
 ds^2= G_{\Lambda\Pi}du^{\Lambda}du^{\Pi},
\end{equation}
with
\begin{equation}
G_{\Lambda\Pi}=E_{\Lambda}^{(\mathbf{A})}E_{\Pi}^{(\mathbf{B})}\eta_{(\mathbf{A})(\mathbf{B})},
\end{equation}
where $ \eta_{(\mathbf{A})(\mathbf{B})}$ and $
E_{\Lambda}^{(\mathbf{A})}$ are flat metric and vielbein
components respectively.
\newline
We choose each
 $ \eta_{(\mathbf{A})(\mathbf{B})}$ as plus or minus  Kronecker's  delta function.
Let us give the 1-form  $\omega^{(\mathbf{A})} $ by
\begin{equation}
\omega^{(\mathbf{A})}= du^{\Lambda} E_{\Lambda}^{(\mathbf{A})}.
\end{equation}
We now define Riemannian normal coordinates by
\begin{equation}
  u^{\Lambda}=v^{\Lambda}t,
\end{equation}
then
\begin{equation}
 du^{\Lambda}=v^{\Lambda}dt+tdv^{\Lambda}.
\end{equation}
Substituting in (2.3)
\begin{equation}
  \omega^{(\mathbf{A})}= tdv^{\Lambda}
  E_{\Lambda}^{(\mathbf{A})}+dtv^{\Lambda}E_{\Lambda}^{(\mathbf{A})}.
\end{equation}
Let us define
\begin{equation}
 z^{(\mathbf{A})}=v^{\Lambda}E_{\Lambda}^{(\mathbf{A})},
\end{equation}
so that
\begin{equation}
\omega^{(\mathbf{A})}=dtz^{(\mathbf{A})}+tdz^{(\mathbf{A})}
+tE^{\Pi(\mathbf{A})}\frac{\partial{E_{\Pi(\mathbf{B})}}}{\partial{z^{(\mathbf{C})}}}z^{(\mathbf{B})}dz^{(\mathbf{C})}.\\
\end{equation}
We now make
\begin{equation}
 A^{(\mathbf{A})_{(\mathbf{B})(\mathbf{C})}}=tE^{\Pi(\mathbf{A})}\frac{\partial{E_{\Pi(\mathbf{B})}}}{\partial{z^{(\mathbf{C})}}},
\end{equation}
then
\begin{equation}
\varpi^{(\mathbf{A})}=
tdz^{(\mathbf{A})}+A^{({A})_{(\mathbf{B})(\mathbf{C})}}z^{(\mathbf{B})}dz^{(\mathbf{C})},
\end{equation}
with
\begin{equation}
\omega^{(\mathbf{A})}=dtz^{(\mathbf{A})}+\varpi^{(\mathbf{A})}.
\end{equation}
We have at $t=0$
\begin{equation}
 A^{({A})_{(\mathbf{B})(\mathbf{C})}}(t=0,z^{(\mathbf{D})})=0,
\end{equation}
\begin{equation}
\varpi^{(\mathbf{A})}(t=0,z^{(\mathbf{D})})=0 ,
\end{equation}
and
\begin{equation}
\omega^{(\mathbf{A})}(t=0,z^{(\mathbf{D})})=dtz^{(\mathbf{A})} .
\end{equation}
We conclude that $\omega^{(\mathbf{A})}$ is the one-form associated
to the normal coordinate $u^{\Lambda}$, $z^{(\mathbf{A})}$ is
associated to the local coordinate $v^{\Lambda}$ of a local basis,
and $\varpi^{(\mathbf{A})}$ is the one-form associated to the
one-form $dz^{(\mathbf{A})}$.
\newline
Consider, at a n+1-manifold, a
coordinate system given by $(t,z^{(\mathbf{A})})$. For each value of
t we have a hyper-surface, where $dt=0$ on each of them. We are
interested in the hyper-surface with $t=1$. On this hyper-surface we
verify the following equality
\begin{equation}
\omega^{(\mathbf{A})}(t=1,z)=\varpi^{(\mathbf{A})}(t=1,z).
\end{equation}
The equality (2.15) is true on all hyper-surface $t=constant.$
\newline
Consider the following expression in a  vielbein basis
\begin{equation}
d\omega^{(\mathbf{A})}=-\omega^{(\mathbf{A})}_{(\mathbf{B})}\wedge\omega^{(\mathbf{B})}.
\end{equation}
The expression is invariant by coordinate transformations.
\newline
Consider now the map  $\Phi$, between two manifolds M and N,
\newline
and consider two subsets, U of M and V of N.Then,
\begin{equation}
\Phi:U\longrightarrow V.
\end{equation}
Define now pull-back as follows, \cite{4},
\begin{equation}
\Phi^\ast:F^p(V)\longrightarrow F^p(U),
\end{equation}
so that $ \Phi^\ast $ sends p-forms into p-forms.
\newline
It is well known that the exterior derivative commutes with
pull-back, so that
\begin{equation}
\Phi^\ast(d\omega^{(\mathbf{A})}_{(\mathbf{B})})=d\Phi^\ast(\omega^{(\mathbf{A})}_{(\mathbf{B})}),
\end{equation}
and
\begin{equation}
\Phi^\ast(d\omega^{(\mathbf{A})})=d\Phi^\ast(\omega^{(\mathbf{A})}).
\end{equation}
We also have
\begin{equation}
\Phi^\ast(\omega^{(\mathbf{A})}_{(\mathbf{B})}\wedge\omega^{(\mathbf{B})})=
\Phi^\ast(\omega^{(\mathbf{A})}_{(\mathbf{B})})\wedge\Phi^\ast(\omega^{(\mathbf{B})}).
\end{equation}
The equation (2.11) can be seen as pull-back,
\begin{equation}
\Phi^\ast(\omega^{(\mathbf{A})})=dtz^{(\mathbf{A})}+\varpi^{(\mathbf{A})}.
\end{equation}
It can be shown, by a simple calculation that
\begin{equation}
\Phi^\ast(\omega^{(\mathbf{A})}_{(\mathbf{B})})=
\varpi^{(\mathbf{A})}_{(\mathbf{B})}.
\end{equation}
We note that $dt=0$, for $\varpi^{(\mathbf{A})}$ and for
$\varpi^{(\mathbf{A})}_{(\mathbf{B})}$.
\newpage
 By the exterior derivative of (2.22), we obtain
\begin{eqnarray}
\nonumber
d(\Phi^\ast(\omega^{(\mathbf{A})}))=d(dtz^{(\mathbf{A})}+\varpi^{(\mathbf{A})})=dz^{(\mathbf{A})}\wedge(dt)\\
\nonumber+dt\wedge\frac{\partial(\varpi^{(\mathbf{A})})}{\partial(t)}\\
\end{eqnarray}
\newline
+ terms not involving  $dt$.
\vspace{1cm}
\newline
Making a pull-back of (2.16) and using (2.21) we have
\begin{equation}
\Phi^\ast(d\omega^{(\mathbf{A})})=\Phi^\ast(-\omega^{(\mathbf{A})}_{(\mathbf{B})}\wedge\omega^{(\mathbf{B})})=
-\Phi^\ast(\omega^{(\mathbf{A})}_{(\mathbf{B})})\wedge\Phi^\ast(\omega^{(\mathbf{B})}).
\end{equation}
Using (2.20), (2.23), (2.24) and (2.25) we have
\begin{equation}
\frac{\partial(\varpi^{(\mathbf{A})})}{\partial(t)}=dz^{(\mathbf{A})}+
\varpi^{(\mathbf{A})}_{(\mathbf{B})}z^{(\mathbf{D})}.
\end{equation}
We can, by a similar procedure to (2.19), and using the  Cartan's
second structure equation, obtain the following result
\begin{equation}
\frac{\partial(\varpi_{(\mathbf{A})(\mathbf{B})})}{\partial(t)}=
R_{(\mathbf{A})(\mathbf{B})(\mathbf{C})(\mathbf{D})}z^{(\mathbf{C})}\varpi^{(\mathbf{A})}.
\end{equation}
Making a new partial derivative of (2.26), two partial derivatives
of (2.10), comparing the results and using (2.27) we have the
following equation
\begin{equation}
\frac{\partial^2(A_{(\mathbf{A}){(\mathbf{C})(\mathbf{D})}})}{\partial(t^2)}=
tz^{(\mathbf{B})}R_{(\mathbf{A})(\mathbf{B})(\mathbf{C})(\mathbf{D})}+
z^{(\mathbf{L})}z^{(\mathbf{M})}R_{(\mathbf{A})(\mathbf{L})(\mathbf{M})(\mathbf{N})}
A_{(\mathbf{P}){(\mathbf{C})(\mathbf{D})}}\eta^{(\mathbf{N})(\mathbf{P})}.
\end{equation}
Rewriting (2.28), with the indices (C) and (D) permuted, we obtain
the following result
\begin{equation}
\frac{\partial^2(A_{(\mathbf{A}){(\mathbf{D})(\mathbf{C})}})}{\partial(t^2)}=
tz^{(\mathbf{B})}R_{(\mathbf{A})(\mathbf{B})(\mathbf{D})(\mathbf{C})}+
z^{(\mathbf{L})}z^{(\mathbf{M})}R_{(\mathbf{A})(\mathbf{L})(\mathbf{M})(\mathbf{N})}
A_{(\mathbf{P}){(\mathbf{D})(\mathbf{C})}}\eta^{(\mathbf{N})(\mathbf{P})}.
\end{equation}
Adding (2.28) and (2.29) and using the curvature symmetries we
have the following solution
\begin{equation}
A_{(\mathbf{A}){(\mathbf{C})(\mathbf{D})}}+A_{(\mathbf{A}){(\mathbf{D})(\mathbf{C})}}=0,
\end{equation}
that is true for all t.
\newline
Then,
\begin{equation}
A_{(\mathbf{A}){(\mathbf{C})(\mathbf{D})}}=-A_{(\mathbf{A}){(\mathbf{D})(\mathbf{C})}},
\end{equation}
so that,  we can rewrite (2.10) as
\begin{equation}
\varpi^{(\mathbf{A})}= tdz^{(\mathbf{A})}+
\frac{1}{2}A^{({A})_{(\mathbf{B})(\mathbf{C})}}(z^{(\mathbf{B})}dz^{(\mathbf{C})}-z^{(\mathbf{C})}dz^{(\mathbf{B})}).
\end{equation}
Let us  define
\begin{equation}
A_{(\mathbf{A}){(\mathbf{C})(\mathbf{D})}}=z^{(\mathbf{B})}B_{(\mathbf{A}){(\mathbf{B})(\mathbf{C})(\mathbf{D})}}.
\end{equation}
The following result is obtained by substituting (2.33) in (2.28),
\begin{equation}
\frac{\partial^2(B_{(\mathbf{A}){(\mathbf{B})(\mathbf{C})(\mathbf{D})}})}{\partial(t^2)}=
tR_{(\mathbf{A})(\mathbf{B})(\mathbf{C})(\mathbf{D})}+
z^{(\mathbf{L})}z^{(\mathbf{M})}R_{(\mathbf{A})(\mathbf{B})(\mathbf{L})(\mathbf{N})}
B_{(\mathbf{P}){(\mathbf{M})(\mathbf{C})(\mathbf{D})}}\eta^{(\mathbf{N})(\mathbf{P})}.
\end{equation}
We now rewrite (2.34) as follows
\begin{equation}
\frac{\partial^2(B_{(\mathbf{B}){(\mathbf{A})(\mathbf{C})(\mathbf{D})}})}{\partial(t^2)}=
tR_{(\mathbf{B})(\mathbf{A})(\mathbf{C})(\mathbf{D})}+
z^{(\mathbf{L})}z^{(\mathbf{M})}R_{(\mathbf{B})(\mathbf{A})(\mathbf{L})(\mathbf{N})}
B_{(\mathbf{P}){(\mathbf{M})(\mathbf{C})(\mathbf{D})}}\eta^{(\mathbf{N})(\mathbf{P})}.
\end{equation}
Adding (2.34) and (2.35) and using the curvature symmetries we
obtain the solution
\begin{equation}
B_{(\mathbf{A}){(\mathbf{B})(\mathbf{C})(\mathbf{D})}}+B_{(\mathbf{B}){(\mathbf{A})(\mathbf{C})(\mathbf{D})}}=const.,
\end{equation}
for all t.
\newline
We can use (2.12) and  (2.33) in (2.36) to obtain
\begin{equation}
B_{(\mathbf{A}){(\mathbf{B})(\mathbf{C})(\mathbf{D})}}+B_{(\mathbf{B}){(\mathbf{A})(\mathbf{C})(\mathbf{D})}}=0.
\end{equation}
In the following, for future use, we present the line element on
the hyper-surface
\begin{equation}
 ds'^2=\eta_{(\mathbf{A})(\mathbf{B})}\varpi^{(\mathbf{A})}\varpi^{(\mathbf{B})}.
\end{equation}
From (2.31), (2.33) and (2.37) we conclude that
$B_{(\mathbf{A}){(\mathbf{B})(\mathbf{C})(\mathbf{D})}} $ has the
same symmetries of the Riemann curvature tensor
\begin{equation}
B_{(\mathbf{A}){(\mathbf{B})(\mathbf{C})(\mathbf{D})}}=-B_{(\mathbf{B}){(\mathbf{A})(\mathbf{C})(\mathbf{D})}}=
-B_{(\mathbf{A}){(\mathbf{B})(\mathbf{D})(\mathbf{C})}}.
\end{equation}
 Using (2.31) and (2.37) we have
\begin{eqnarray}
\nonumber A_{(\mathbf{A}){(\mathbf{C})(\mathbf{D})}}dz^{(\mathbf{A})}z^{(\mathbf{C})}dz^{(\mathbf{D})}=\\
\nonumber +\frac{1}{4}B_{(\mathbf{A}){(\mathbf{B})(\mathbf{C})(\mathbf{D})}}.\\
\nonumber .(z^{(\mathbf{B})}dz^{(\mathbf{A})}-z^{(\mathbf{A})}dz^{(\mathbf{B})}).\\
\nonumber .(z^{(\mathbf{C})}dz^{(\mathbf{D})}-z^{(\mathbf{D})}dz^{(\mathbf{C})}).\\
\end{eqnarray}
Now we can construct the line element of the hyper-surface. By
direct use of (2.32) and (2.40) we have
\begin{eqnarray}
\nonumber ds'^2=t^2\eta_{(\mathbf{A})(\mathbf{B})}dz^{(\mathbf{A})}dz^{(\mathbf{B})}+\\
 \nonumber+\frac{1}{2}\{\frac{1}{2}t\epsilon_{(\mathbf{B})}B_{(\mathbf{A}){(\mathbf{B})(\mathbf{C})(\mathbf{D})}}+\\
 \nonumber+\eta^{(\mathbf{M})(\mathbf{N})}A_{(\mathbf{M}){(\mathbf{B})(\mathbf{A})}}A_{(\mathbf{N}){(\mathbf{C})(\mathbf{D})}}\}.\\
\nonumber.(z^{(\mathbf{B})}dz^{(\mathbf{A})}-z^{(\mathbf{A})}dz^{(\mathbf{B})})(z^{(\mathbf{C})}dz^{(\mathbf{D})}-z^{(\mathbf{D})}dz^{(\mathbf{C})}).\\
\end{eqnarray}
The line elements of the manifold and the hyper-surface are equal
at $ t=1 $, where  $u^{\Lambda}=v^{\Lambda} $,
\begin{equation}
ds^2=ds'^2,
\end{equation}
and
\begin{eqnarray}
\nonumber ds^2=\eta_{(\mathbf{A})(\mathbf{B})}dz^{(\mathbf{A})}dz^{(\mathbf{B})}+\\
 \nonumber+\frac{1}{2}\{\frac{1}{2}\epsilon_{(\mathbf{B})}B_{(\mathbf{A}){(\mathbf{B})(\mathbf{C})(\mathbf{D})}}+\\
 \nonumber+\eta^{(\mathbf{M})(\mathbf{N})}A_{(\mathbf{M}){(\mathbf{B})(\mathbf{A})}}A_{(\mathbf{N}){(\mathbf{C})(\mathbf{D})}}\}.\\
\nonumber.(z^{(\mathbf{B})}dz^{(\mathbf{A})}-z^{(\mathbf{A})}dz^{(\mathbf{B})})(z^{(\mathbf{C})}dz^{(\mathbf{D})}-z^{(\mathbf{D})}dz^{(\mathbf{C})}).\\
\end{eqnarray}
Note that (2.43) is not an approximation of (2.1), they are equal.
\newpage
\renewcommand{\theequation}{\thesection.\arabic{equation}}
\section{\bf Conformal Form of Riemannian Metrics }
 \setcounter{equation}{0}
$         $
 \setcounter{equation}{0}
 $         $
Sometimes it is possible to write the metric in a particular form,
as follows
\begin{eqnarray}
\nonumber ds^2=\eta_{(\mathbf{a})(\mathbf{b})}dz^{(\mathbf{a})}dz^{(\mathbf{b})}+\\
 \nonumber+\{\eta_{(\mathbf{0})(\mathbf{0})}+\frac{1}{2}[\frac{1}{2}\epsilon_{(\mathbf{B})}B_{(\mathbf{A}){(\mathbf{B})(\mathbf{C})(\mathbf{D})}}+\\
 \nonumber+\eta^{(\mathbf{M})(\mathbf{N})}A_{(\mathbf{M}){(\mathbf{B})(\mathbf{A})}}A_{(\mathbf{N}){(\mathbf{C})(\mathbf{D})}}].\\
\nonumber.(z^{(\mathbf{B})}\frac{dz^{(\mathbf{A})}}{d\tau}-z^{(\mathbf{A})}\frac{dz^{(\mathbf{B})}}{d\tau})(z^{(\mathbf{C})}\frac{dz^{(\mathbf{D})}}{d\tau}-z^{(\mathbf{D})}\frac{dz^{(\mathbf{C})}}{d\tau})\}d\tau^2,\\
\end{eqnarray}
where$(a),(b)\neq 0 $.
 \newline
 Defining
\begin{eqnarray}
d\rho^2=\{\eta_{(\mathbf{0})(\mathbf{0})}+\frac{1}{2}[\frac{1}{2}\epsilon_{(\mathbf{B})}B_{(\mathbf{A}){(\mathbf{B})(\mathbf{C})(\mathbf{D})}}+\\
 \nonumber+\eta^{(\mathbf{M})(\mathbf{N})}A_{(\mathbf{M}){(\mathbf{B})(\mathbf{A})}}A_{(\mathbf{N}){(\mathbf{C})(\mathbf{D})}}].\\
\nonumber.(z^{(\mathbf{B})}\frac{dz^{(\mathbf{A})}}{d\tau}-z^{(\mathbf{A})}\frac{dz^{(\mathbf{B})}}{d\tau})(z^{(\mathbf{C})}\frac{dz^{(\mathbf{D})}}{d\tau}-z^{(\mathbf{D})}\frac{dz^{(\mathbf{C})}}{d\tau})\}d\tau^2,\\
\end{eqnarray}
then, (3.1) can be rewritten as
\begin{equation}
ds^2=d\rho^2+\eta_{(\mathbf{a})(\mathbf{b})}dz^{(\mathbf{a})}dz^{(\mathbf{b})}.
\end{equation}
We now write (2.43) as
\begin{eqnarray}
 \nonumber ds^2=\eta_{(\mathbf{A})(\mathbf{B})}dz^{(\mathbf{A})}dz^{(\mathbf{B})}+\\
 \nonumber +\{\frac{1}{2}[\frac{1}{2}\epsilon_{(\mathbf{B})}B_{(\mathbf{A}){(\mathbf{B})(\mathbf{C})(\mathbf{D})}}+\\
 \nonumber +\eta^{(\mathbf{M})(\mathbf{N})}A_{(\mathbf{M}){(\mathbf{B})(\mathbf{A})}}A_{(\mathbf{N}){(\mathbf{C})(\mathbf{D})}}]\}.\\
\nonumber .(z^{(\mathbf{B})}\frac{dz^{(\mathbf{A})}}{ds}-z^{(\mathbf{A})}\frac{dz^{(\mathbf{B})}}{ds})(z^{(\mathbf{C})}\frac{dz^{(\mathbf{D})}}{ds}-z^{(\mathbf{D})}\frac{dz^{(\mathbf{C})}}{ds}))ds^2.\\
\end{eqnarray}
\newpage
It can also be written in the form
\begin{eqnarray}
 \nonumber[1-\frac{1}{2}[\frac{1}{2}\epsilon_{(\mathbf{B})}B_{(\mathbf{A}){(\mathbf{B})(\mathbf{C})(\mathbf{D})}}+\\
 \nonumber+\eta^{(\mathbf{M})(\mathbf{N})}A_{(\mathbf{M}){(\mathbf{B})(\mathbf{A})}}A_{(\mathbf{N}){(\mathbf{C})(\mathbf{D})}}].\\
\nonumber.(z^{(\mathbf{B})}\frac{dz^{(\mathbf{A})}}{ds}-z^{(\mathbf{A})}\frac{dz^{(\mathbf{B})}}{ds})(z^{(\mathbf{C})}\frac{dz^{(\mathbf{D})}}{ds}-z^{(\mathbf{D})}\frac{dz^{(\mathbf{C})}}{ds})]ds^2\\
 \nonumber =\eta_{(\mathbf{A})(\mathbf{B})}dz^{(\mathbf{A})}dz^{(\mathbf{B})}.\\
\end{eqnarray}
We now define the function
\begin{eqnarray}
L^{\mathbf{A})(\mathbf{B})}=(z^{(\mathbf{B})}\frac{dz^{(\mathbf{A})}}{ds}-z^{(\mathbf{A})}\frac{dz^{(\mathbf{B})}}{ds}),
\end{eqnarray}
which is the classical angular momentum  of a free particle.
\newline
The line element (3.6) can assume the following form
\begin{eqnarray}
 \nonumber\{1+\frac{1}{2}[\frac{1}{2}(\epsilon_{(\mathbf{B})}B_{(\mathbf{A}){(\mathbf{B})(\mathbf{C})(\mathbf{D})}}+\\
 \nonumber+\eta^{(\mathbf{M})(\mathbf{N})}A_{(\mathbf{M}){(\mathbf{B})(\mathbf{A})}}A_{(\mathbf{N}){(\mathbf{C})(\mathbf{D})}}].\\
\nonumber.(L^{\mathbf{A})(\mathbf{B})}L^{\mathbf{C})(\mathbf{D})})\}ds^2\\
 \nonumber =(\eta_{(\mathbf{A})(\mathbf{B})}dz^{(\mathbf{A})}dz^{(\mathbf{B})}.\\
\end{eqnarray}
We now define the function
\begin{eqnarray}
\nonumber \exp(-2\sigma)=\{1+\frac{1}{2}[\frac{1}{2}(\epsilon_{(\mathbf{B})}B_{(\mathbf{A}){(\mathbf{B})(\mathbf{C})(\mathbf{D})}}\\
 \nonumber +\eta^{(\mathbf{M})(\mathbf{N})}A_{(\mathbf{M}){(\mathbf{B})(\mathbf{A})}}A_{(\mathbf{N})){(\mathbf{C})(\mathbf{D})}})].\\
\nonumber .L^{(\mathbf{A})(\mathbf{B})}L^{(\mathbf{C})(\mathbf{D})}\},\\
\end{eqnarray}
so that, the line element assumes the form
\begin{equation}
ds^2=\exp(2\sigma)\eta_{(\mathbf{A})(\mathbf{B})}dz^{(\mathbf{A})}dz^{(\mathbf{B})}.
\end{equation}
\newpage
When  transformations like (3.2) are possible, (3.4) will be a flat
metric, with the time changed, and it is equivalent to the original
metric. The metric (3.10) is conformal to a flat manifold, and we
conclude that all n-dimensional pseudo-Riemannian metrics are
conformal to  flat manifolds, when, in normal coordinates, the
transformations are well-behaved in the origin and in its
neighborhood. It is important to pay attention to the fact that a
normal transformation and its inverse are well-behaved in the region
where geodesics  are not mixed. Points where geodesics close or mix
are known as conjugate points of Jacobi's fields. Jacobi's fields
can be used for this purpose. Although this is an important problem,
we do not make other considerations about the regions where (3.4)
and (3.10) will be valid.
\newline
In the next section we present the Cartan's solution for the case
where curvature is constant. For the Cartan's solution to a general
metric, more geometric objects, like normal tensors, are necessary.
This will be presented in section $5$.
\section{Cartan's Solution for Constant Curvature}
  $                      $
 \setcounter{equation}{0}
 $         $
In this section we present the Cartan's solution for the constant
curvature. The calculation is very simple and was done in \cite{3},
and reproduced in  detail in \cite{4}. Our objective in this section
is only to place the Cartan's solution in the forms (3.4) and
(3.10).
\newline
Cartan used the signature $(+,+,+....,+)$ and obtained the
following line element
\begin{eqnarray}
 \nonumber
 ds^2=\sum_{k=1}^{n}(\varpi^{\mathbf{k}})^2=\sum_{k=1}^{n}(dv^{\mathbf{k}})^2+\\
 \nonumber -[\frac{|K|{\mathbf{r^2}} -{\mathbf{S^2(r\sqrt{|K|}t)}}}{|K|\mathbf{r^4}}]\sum_{i<j}(v^{\mathbf{i}}dv^{\mathbf{j}}-v^{\mathbf{j}}dv^{\mathbf{i}})^2,\\
\end{eqnarray}
where for $K>0$
\begin{equation}
  \mathbf{S}=\sin( \sqrt{|K|}t),
\end{equation}
 and for $K<0$
\begin{equation}
  \mathbf{S}=\sinh( \sqrt{|K|}t).
\end{equation}
 We write (4.1) in the form (3.1)
\begin{eqnarray}
 \nonumber
 ds^2=\sum_{k=1}^{n}(\varpi^{\mathbf{k}})^2=\sum_{k=1}^{n}(dv^{\mathbf{k}})^2\\
 \nonumber -[\frac{|K|{\mathbf{r^2}} -{\mathbf{S^2(r\sqrt{|K|}t)}}}{|K|\mathbf{r^4}}]\sum_{i<j}(v^{\mathbf{i}}\frac{dv^{\mathbf{j}}}{d\tau}-v^{\mathbf{j}}\frac{dv^{\mathbf{i}}}{d\tau})^2d\tau^2.\\
\end{eqnarray}
 Consider the following function
\begin{equation}
l^{\mathbf{ij}}=\sum_{i<j}(v^{\mathbf{i}}\frac{dv^{\mathbf{j}}}{d\tau}-v^{\mathbf{j}}\frac{dv^{\mathbf{i}}}{d\tau})^2.
\end{equation}
\newpage
Using (4.5) in (4.4) we obtain
\begin{eqnarray}
\nonumber
 ds^2=\sum_{k=1}^{n}(\varpi^{\mathbf{k}})^2=\sum_{k=1}^{n}(dv^{\mathbf{k}})^2\\
 \nonumber -[\frac{|K|{\mathbf{r^2}} -{\mathbf{S^2(r\sqrt{|K|}t)}}}{|K|\mathbf{r^4}}]\sum_{i<j}(l^{\mathbf{ij}})^2d\tau^2.\\
\end{eqnarray}
Sometimes we can suppose that $ dv^1=d\tau $. Then, in this case
(4.6) can be written in the form
\begin{eqnarray}
 \nonumber ds^2=\sum_{k=1}^{n}(\varpi^{\mathbf{k}})^2=\sum_{k=2}^{n}(dv^{\mathbf{k}})^2\\
 \nonumber +\{1 -[\frac{|K|{\mathbf{r^2}} -{\mathbf{S^2(r\sqrt{|K|}t)}}}{|K|\mathbf{r^4}}]\sum_{i<j}(l^{\mathbf{ij}})^2\}d\tau^2.\\
\end{eqnarray}
Defining,
\begin{equation}
 d\rho^2=\{1-[\frac{|K|{\mathbf{r^2}}-{\mathbf{S^2(r\sqrt{|K|}t)}}}{|K|\mathbf{r^4}}]\sum_{i<j}(l^{\mathbf{ij}})^2\}d\tau^2\\,
\end{equation}
and using it in (4.6), we obtain
\begin{equation}
ds^2=d\rho^2 +\sum_{k=2}^{n}(dv^{\mathbf{k}})^2,
\end{equation}
where (4.9) has the same form as (3.4).
\newline
We now write (4.1) in the form (3.10). For this we change (4.1) as
follows
\begin{eqnarray}
 \nonumber
 ds^2=\sum_{k=1}^{n}(\varpi^{\mathbf{k}})^2=\sum_{k=1}^{n}(dv^{\mathbf{k}})^2+\\
 \nonumber -[\frac{|K|{\mathbf{r^2}} -{\mathbf{S^2(r\sqrt{|K|}t)}}}{|K|\mathbf{r^4}}]\sum_{i<j}(v^{\mathbf{i}}\frac{dv^{\mathbf{j}}}{ds}-v^{\mathbf{j}}\frac{dv^{\mathbf{i}}}{ds})^2ds^2.\\
\end{eqnarray}
We note that (4.10) has the form of (3.5).
\newline
Defining
\begin{eqnarray}
L^{(\mathbf{i})(\mathbf{j})}=(z^{(\mathbf{i})}\frac{dz^{(\mathbf{j})}}{ds}-z^{(\mathbf{j})}\frac{dz^{(\mathbf{i})}}{ds}).
\end{eqnarray}
and replacing (4.11) in ( 4.10) we obtain
 \begin{eqnarray}
 \nonumber
 ds^2=\sum_{k=1}^{n}(\varpi^{\mathbf{k}})^2=\sum_{k=1}^{n}(dv^{\mathbf{k}})^2+\\
 \nonumber -[\frac{|K|{\mathbf{r^2}} -{\mathbf{S^2(r\sqrt{|K|}t)}}}{|K|\mathbf{r^4}}]\sum_{i<j}(L^{(\mathbf{i})(\mathbf{j})})^2ds^2,\\
\end{eqnarray}
which is equivalent to
\begin{eqnarray}
\{1+[\frac{|K|{\mathbf{r^2}}
-{\mathbf{S^2(r\sqrt{|K|}t)}}}{|K|\mathbf{r^4}}]\sum_{i<j}(L^{(\mathbf{i})(\mathbf{j})})^2\}ds^2=\sum_{k=1}^{n}(dv^{\mathbf{k}})^2.
\end{eqnarray}
We now define
\begin{eqnarray}
\exp(-2\sigma)=\{1+[\frac{|K|{\mathbf{r^2}}
-{\mathbf{S^2(r\sqrt{|K|}t)}}}{|K|\mathbf{r^4}}]\sum_{i<j}(L^{(\mathbf{i})(\mathbf{j})})^2\}.
\end{eqnarray}
Substituting in (4.13) we obtain
\begin{equation}
ds^2=\exp(2\sigma)\sum_{k=1}^{n}(dv^{\mathbf{k}})^2.
\end{equation}
We could have this section with all  equations in a vielbein basis.
The results would be the same. This will be made at the end of the
next section for the general solution.
\newline
We rewrite (4.15) as follows
\begin{eqnarray}
ds^2=\{1+[\frac{|K|{\mathbf{r^2}}
-{\mathbf{S^2(r\sqrt{|K|}t)}}}{|K|\mathbf{r^4}}]\sum_{i<j}(\eta_{\mathbf{i}\mathbf{j}}L^{(\mathbf{i})(\mathbf{j})})^2\}^{-1}dv^{\mathbf{l}}dv^{\mathbf{k}}\eta_{\mathbf{l}\mathbf{k}},
\end{eqnarray}
where  $\eta_{\mathbf{j}\mathbf{k}}$ is a generic flat metric.
\newline
By a coordinate transformation we  can put (4.16) in the well known
form
\begin{eqnarray}
ds^2=\{1+\frac{K\Omega^{\mathbf{j}}\Omega^{\mathbf{k}}\eta_{\mathbf{j}{\mathbf{k}}}}{4}\}^{-2}d\Omega^{\mathbf{j}}d\Omega^{\mathbf{k}}\eta_{\mathbf{j}{\mathbf{k}}}.
\end{eqnarray}
It is well known that (4.17) is conformal to a flat metric. As
(4.16) and (4.17) are equivalent, we conclude that (4.16) is also
conformal to a flat metric. Therefore, we conclude that there is a
local conformal transformation between (4.16) or (4.17) and (3.10).
This is an important result that will be analyzed in section 7.
\newline
In the next section we present, in detail, some geometric objects,
like normal tensors. This is necessary for the Cartan's solution of
a general metric.
\newline
\section{Normal Tensors}
  $                      $
 \setcounter{equation}{0}
 $         $
In this section, a Taylor's expansion for the metric tensor
components will be built in the origin of  normal coordinates.
Normal tensors are very important for this. In this paper we use the
notation (;) for the covariant derivative.
\newline
Consider the line element
\begin{equation}
 ds^2= G_{\Lambda\Pi}du^{\Lambda}du^{\Pi}.
\end{equation}
Its expansion in the origin of a normal coordinate has the general
form
\begin{eqnarray}
 \nonumber ds^2=G_{\lambda\pi}du^{\lambda}du^{\pi}=G_{\lambda\pi}(0)+\frac{\partial{G_{\lambda\pi}}}{\partial{u^{\mu}}}v^{\mu}t\\
\nonumber+\frac{1}{2}\frac{\partial^2G_{\lambda\pi}}{\partial{u^{\mu}}\partial{u^{\nu}}}v^{\mu}v^{\nu}t^2+.......,\\
\end{eqnarray}
where the derivatives are calculated at $ u^{\pi}=0 $.
\newline
Some results will be found in \cite{6}, \cite{7}, but, in general,
they are not simple. Our results are simpler because they are more
specific.
\newline
Consider the covariant derivative of $G_{\lambda\pi}$ at a normal
coordinate.
\newline
For a pseudo-Riemannian space we have
\begin{equation}
 G_{\lambda\pi};_{\mu}=0.
\end{equation}
From (5.3) we obtain
\begin{equation}
 \frac{\partial{G_{\lambda\pi}}}{\partial{u^{\mu}}}=C^{\rho}_{\mu\lambda}G_{\rho\pi}
+C^{\rho}_{\mu\pi}G_{\lambda\rho},
\end{equation}
where
\begin{equation}
C^{\rho}_{\mu\lambda}(0)=0,
\end{equation}
and
\begin{equation}
 \frac{\partial{G_{\lambda\pi}}}{\partial{u^{\mu}}}(0)=0,
\end{equation}
in origin.
\newline
In the  limit  $u=0$, the  partial derivatives of (5.4) supply all
derivative terms for the expansion (5.2). Each partial derivative of
$C^{\rho}_{\mu\lambda}$, calculated in the origin, is a new tensor.
These new tensors are called normal tensors. We designate the
following representation for them,
\begin{equation}
D^{\rho}_{\mu\lambda\alpha\beta....\gamma}=
\frac{\partial^nC^{\rho}_{\mu\lambda}}{\partial{u^{\alpha}}\partial{u^{\beta}}...\partial{u^{\gamma}}}(0).
\end{equation}
We conclude from (5.7) that normal tensors are symmetric at the
first pair of inferior indices and also have  a complete symmetry
among other inferior indices.
\newline
It is simple to show that
\begin{equation}
S(D^{\rho}_{\mu\lambda\alpha\beta....\gamma})=0,
\end{equation}
where S designates the sum of different normal tensor components.
With (5.4),(5.5), (5.6), (5.7) and (5.8) we can calculate all terms
of the expansion (5.2).
\newline
Deriving (5.5), calculating the limit, and using (5.7) we have
\begin{equation}
 \frac{\partial^2{G_{\lambda\pi}}}{\partial{u^{\mu}}{u^{\nu}}}=G_{\lambda\rho}D^{\rho}_{\mu\pi\nu}
+G_{\pi\rho}D^{\rho}_{\mu\lambda\nu}.
\end{equation}
There is more than one way of associating  the curvature tensor with
normal tensors. In the following we present the simplest way we
know.
\newline
Let us define, in normal coordinates, the following components of
the Riemannian curvature tensor
\begin{equation}
R^{\rho}_{\mu\lambda\nu}=\frac{\partial(C^{\rho}_{\mu\lambda})}{\partial{u^{\nu}}}-\frac{\partial(C^{\rho}_{\mu\nu})}{\partial{u^{\lambda}}}
+C^{\sigma}_{\mu\lambda}C^{\rho}_{\sigma\nu}-C^{\sigma}_{\mu\nu}C^{\rho}_{\sigma\lambda}.
\end{equation}
The limit of (5.10) is
\begin{equation}
R^{\rho}_{\mu\lambda\nu}=D^{\rho}_{\mu\lambda\nu}-D^{\rho}_{\mu\nu\lambda},
\end{equation}
where we have used (5.5) and (5.7).
\newline
Using (5.7), (5.8),  (5.11) and the symmetries of the Riemannian
curvature tensor, we can  show that
\begin{equation}
D^{\rho}_{\mu\lambda\nu}=\frac{1}{3}(R^{\rho}_{\mu\lambda\nu}+R^{\rho}_{\lambda\mu\nu}).
\end{equation}
Using (5.9) and (5.12) we obtain
\begin{equation}
 \frac{\partial^2{G_{\alpha\beta}}}{\partial{u^{\gamma}}{u^{\delta}}}u^{\gamma}u^{\delta}=\frac{2}{3}R_{\alpha\gamma\beta\delta}u^{\gamma}u^{\delta}.
\end{equation}
By similar procedure, but tedious calculation, we obtain
\begin{equation}
 \frac{\partial^3{G_{\alpha\beta}}}{\partial{u^{\mu}}{u^{\nu}}{u^{\sigma}}}u^{\mu}u^{\nu}u^{\sigma}
 =R_{\alpha\mu\beta\nu;\sigma}u^{\gamma}u^{\delta}u^{\mu}u^{\nu}u^{\sigma}.
\end{equation}
Derivatives of fourth order for metric tensor are easy but very
long. We do not present them here.
\newline
Now we can  conclude the Taylor's expansion of the metric tensor.
First we rewrite
\begin{eqnarray}
 \nonumber G_{\lambda\pi}=G_{\lambda\pi}(0)\\
 \nonumber +\frac{1}{2}\frac{\partial^2G_{\lambda\pi}}{\partial{u^{\mu}}\partial{u^{\nu}}}v^{\mu}v^{\nu}t^2\\
 \nonumber \frac{1}{6}\frac{\partial^3{G_{\alpha\beta}}}{\partial{u^{\mu}}{u^{\nu}}{v^{\sigma}}}v^{\mu}v^{\nu}v^{\sigma}t^3+...,\\
\end{eqnarray}
 Now we substitute (5.13) and  (5.14) in (5.15) obtaining
 \begin{eqnarray}
 \nonumber G_{\lambda\pi}du^{\alpha}du^{\beta}=G_{\alpha\beta}(0)du^{\alpha}du^{\beta}+\\
 \nonumber +\frac{1}{3}[R_{\alpha\gamma\beta\delta}t^2+\\
 \nonumber+ \frac{1}{2}v^{\sigma}R_{\alpha\mu\beta\nu;\sigma}t^3+...]v^{\gamma}v^{\delta}du^{\alpha}du^{\beta},.\\
\end{eqnarray}
Using the symmetries of the curvature tensor we have the following
expansion
\begin{eqnarray}
 \nonumber G_{\lambda\pi}du^{\alpha}du^{\beta}=G_{\alpha\beta}(0)du^{\alpha}du^{\beta}+\\
 \nonumber +\frac{1}{12}[R_{\alpha\gamma\beta\delta}t^2+\\
 \nonumber+ \frac{1}{2}v^{\sigma}R_{\alpha\gamma\beta\delta;\sigma}t^3+...][v^{\gamma}du^{\alpha}-v^{\alpha}du^{\gamma}][v^{\beta}du^{\delta}-v^{\delta}du^{\beta}].\\
\end{eqnarray}
On the hyper-surface $t=1$ we have $dt=0$ and
\begin{eqnarray}
 \nonumber G_{\lambda\pi}du^{\alpha}du^{\beta}=G_{\alpha\beta}(0)dv^{\alpha}dv^{\beta}+\\
 \nonumber +\frac{1}{12}[R_{\alpha\gamma\beta\delta}+\\
 \nonumber+ \frac{1}{2}v^{\sigma}R_{\alpha\gamma\beta\delta;\sigma}][v^{\gamma}dv^{\alpha}-v^{\alpha}dv^{\gamma}][v^{\beta}dv^{\delta}-v^{\delta}dv^{\beta}].\\
\end{eqnarray}
which is the same result of Cartan, although, by a different way.
\newline
It is always possible to place a flat metric into a diagonal form.
This is the case of a metric at the origin of normal coordinates. In
this case we have
\begin{equation}
E_{\Lambda}^{(\mathbf{A})}(0)=\delta_{\Lambda}^{(\mathbf{A})}.
\end{equation}
We now present the Taylor's expansion of
 $E_{\Lambda}^{(\mathbf{A})}$ at the origin of a normal
 coordinate
\begin{eqnarray}
\nonumber E_{\Lambda}^{(\mathbf{A})}(u)=\delta_{\Lambda}^{(\mathbf{A})}+\\
 \nonumber+\frac{\partial(E_{\Lambda}^{(\mathbf{A})})}{\partial(u^{\alpha})}du^{\alpha}+...\\
\end{eqnarray}
\newpage
Multiplying (5.18) by the vielbein components and their
inverse, using (5.19) and (5.20), we obtain
\begin{eqnarray}
\nonumber ds^2=\eta_{(\mathbf{A})(\mathbf{B})}dz^{(\mathbf{A})}dz^{(\mathbf{B})}+\\
 \nonumber+\frac{1}{12}[R_{(\mathbf{A}){(\mathbf{B})(\mathbf{C})(\mathbf{D})}}+\\
 \nonumber+\frac{1}{2}z^{(\mathbf{M})}R_{(\mathbf{A}){(\mathbf{B})(\mathbf{C})(\mathbf{D}),(\mathbf{M})}}].\\
\nonumber.(z^{(\mathbf{B})}dz^{(\mathbf{A})}-z^{(\mathbf{A})}dz^{(\mathbf{B})})(z^{(\mathbf{C})}dz^{(\mathbf{D})}-z^{(\mathbf{D})}dz^{(\mathbf{C})}),\\
\end{eqnarray}
where the calculation was made on the hyper-surface $t=1$ and
$dt=0$.
\newline
Note that the expansion given by (5.22) is an approximated solution
of (2.43). Using a perturbation method, Cartan  first solved  the
equations (2.28) and (2.34) and then placed each solution into
(2.43).
\newline
Following the same procedure used to place (5.18) in the form
(5.21), we can place (3.10) as follows
\begin{eqnarray}
\nonumber ds^2=\{1+\frac{1}{2}[\frac{1}{2}(\epsilon_{\beta}B_{\alpha{\beta\gamma\delta}})\\
 \nonumber +\eta^{\rho\sigma}A_{\rho{\alpha\beta}}A_{\sigma{\gamma\delta)}})].\\
\nonumber .L^{\alpha\beta}L^{\gamma\delta}\}^{-1}\eta_{\alpha\beta}d\Omega^{\alpha}d\Omega^{\beta}.\\
\end{eqnarray}
We now rewrite (4.17) obtaining
\begin{eqnarray}
ds'^2=\{1+\frac{K\Omega^{\mathbf{\alpha}}\Omega^{\beta}\eta_{\alpha\beta}}{4}\}^{-2}d\Omega^{\rho}d\Omega^{\sigma}\eta_{\varrho\sigma}.
\end{eqnarray}
Because (5.22) and (5.23) are conformal to a flat manifold, there is
a conformal transformation between them, with a conformal factor,
$(\exp2\psi)$. Then
\begin{eqnarray}
g'_{\alpha\beta}=(\exp2\psi)g_{\alpha\beta}.
\end{eqnarray}
More specifically,
\begin{eqnarray}
\nonumber\{1+\frac{1}{2}[\frac{1}{2}(\epsilon_{\beta}B_{\alpha{\beta\gamma\delta}})+\\
\nonumber
+\eta^{\rho\sigma}A_{\rho{\alpha\beta}}A_{\sigma{\gamma\delta)}})]L^{\alpha\beta}L^{\gamma\delta}\}=\\
 \nonumber
 =(\exp2\psi)\{1+\frac{K\Omega^{\mathbf{\alpha}}\Omega^{\beta}\eta_{\alpha\beta}}{4}\}^{2}.\\
 \end{eqnarray}
\newline
This is an important result with some consequences as we will see.
\newline
Note that (5.23) is a particular Einstein's space with a constant
curvature, where
\begin{eqnarray}
R'_{\alpha\beta}=\frac{R'}{n}g'_{\alpha\beta},
\end{eqnarray}
and $R'$ is the scalar curvature. Spaces, as the Schwarzschild's,
where
\begin{eqnarray}
R_{\alpha\beta}=0,
\end{eqnarray}
are  Einstein's spaces and are not maximally symmetric.
\newline
Einstein's spaces with a constant scalar curvature obey homogeneity
and isotropy conditions. They are maximally symmetric spaces.
\newline
We will be using the following definitions, \cite{8}
\begin{eqnarray}
\triangle_{1}{\psi}=g^{\mu\nu}{\psi}_{,\mu}{\psi}_{,\nu},
\end{eqnarray}
\begin{eqnarray}
{\psi}_{\mu\nu}={\psi}_{;\mu\nu}-{\psi}_{,\mu}{\psi}_{,\nu},
\end{eqnarray}
\begin{eqnarray}
\triangle_{2}{\psi}=g^{\mu\nu}{\psi}_{;\mu\nu}.
\end{eqnarray}
From (5.24), (5.28), (5.29), and (5.30) we obtain
\begin{eqnarray}
\nonumber {\psi}_{\mu\nu}=\frac{1}{(n-2)}(R_{\mu\nu})\\
\nonumber -\frac{1}{(2)(n-1)(n-2)}(g'_{\mu\nu}R'-g_{\mu\nu}R)\\
\nonumber-\frac{1}{2}\triangle_{1}{\psi}g_{\mu\nu}.\\
\end{eqnarray}
If $g'_{\mu\nu}$ is a metric of an Einstein's space, then (5.31) is
simplified to
\begin{eqnarray}
\nonumber {\psi}_{\mu\nu}=-\frac{1}{(n-2)}R_{\mu\nu}+\\
\nonumber +(\frac{1}{(2)(n-1)(n-2)}R+\frac{1}{(2n)(n-1)}R'(\exp2\psi)-\frac{1}{2}\triangle_{1}{\psi})g_{\mu\nu}.\\
\end{eqnarray}
In the region where (2.4) is well-behaved, (5.25) will be possible.
\section{Local Embedding of Riemannian Manifolds in Flat Manifolds}
 $        $
 \setcounter{equation}{0}
 $         $
In section 3 we presented some considerations about the regions
where coordinate transformations are well-defined. We consider that
this is the case, where  such conditions are satisfied.
\newline
Let us rewrite (3.10)
\begin{equation}
ds^2=\exp(2\sigma)\eta_{(\mathbf{A})(\mathbf{B})}dz^{(\mathbf{A})}dz^{(\mathbf{B})}.
\end{equation}
Defining the following  transformation of coordinates, \cite{8},
\begin{equation}
y^{(\mathbf{A})}=\exp(\sigma)z^{(\mathbf{A})},
\end{equation}
with $ (A)=(1,2,3,....,n),$
\begin{equation}
y^{n+1}=\exp(\sigma)(\eta_{(\mathbf{A})(\mathbf{B})}z^{(\mathbf{A})}z^{(\mathbf{B})}-\frac{1}{4}),
\end{equation}
and,
\begin{equation}
y^{n+2}=\exp(\sigma)(\eta_{(\mathbf{A})(\mathbf{B})}z^{(\mathbf{A})}z^{(\mathbf{B})}+\frac{1}{4}).
\end{equation}
It is easy to verify that
\begin{equation}
\eta_{\mathbf{A}\mathbf{B}}y^{\mathbf{A}}y^{\mathbf{B}}=0,
\end{equation}
where,
\begin{equation}
\eta_{\mathbf{A}\mathbf{B}}=(\eta_{(\mathbf{A})(\mathbf{B})},\eta_{\mathbf{(n+1),}\mathbf{(n+1)}},\eta_{\mathbf{(n+2),}\mathbf{(n+2)}}),
\end{equation}
with,
\begin{equation}
\eta_{\mathbf{(n+1),}\mathbf{(n+1)}}=1,
\end{equation}
and,
\begin{equation}
\eta_{\mathbf{(n+2),}\mathbf{(n+2)}=-1}.
\end{equation}
By a simple calculation we can verify that the line elements are
given by
\begin{equation}
ds^2=\exp(2\sigma)\eta_{(\mathbf{A})(\mathbf{B})}dz^{(\mathbf{A})}dz^{(\mathbf{B})}=
\eta_{\mathbf{A}\mathbf{B}}dy^{\mathbf{A}}dy^{\mathbf{B}}.
\end{equation}
The equation (6.5) is a hyper-cone in the (n+2)-dimensional flat
manifold. The metric (6.1) was embedded in the hyper-cone (6.5) of
the (n+2)-dimensional flat manifold. We could present more results
about embedding, but we have already reached our objective.
\newpage
\section{Embedding of Manifolds of  Constant Curvatures in Flat Manifolds}
 $                       $
 \setcounter{equation}{0}
 $         $
In this section we embed  the n-dimensional manifold (5.23) in a
n+1-dimensional flat manifold obtaining, as a geometric result,
without postulate, the quantum angular momentum of a particle. Other
results will be presented in another section.
\newline
We now consider a manifold (5.23) designated by S, embedded in a
n+1-dimensional flat manifold. The following constraint is obeyed
\cite{9},
\begin{eqnarray}
\eta_{\alpha\beta}x^{\alpha}x^{\beta}=K=\epsilon\frac{1}{R^2},
\end{eqnarray}
where K is the  scalar curvature of the n-dimensional manifold
(5.23), $\alpha, \beta = ( 1, 2,...,n+1 )$ and $ \epsilon=(+1, -1).$
For the special case of a n-sphere we use the following notation
$S^n$ for (5.23).
\newline
It is convenient that we use a local basis
$X_{\beta}=\frac{\partial}{\partial(x^{\beta})}.$
\newline
We consider a constant vector $\textbf{C}$ in the n+1-dimensional
manifold given by
\begin{eqnarray}
\eta_{\alpha\beta}C^{\alpha}X^{\beta}=\eta^{\alpha\beta}C_{\alpha}X_{\beta}=C,
\end{eqnarray}
where  $C^{\alpha}$ are constant and  $\textbf{N}$ is a unitary and
normal vector to S. We use the symbol $<,>$ for the internal product
in the n+1-dimensional flat manifold  and $<,>'$ for S.
\newline
A constant vector $\textbf{C}$ can be  decomposed into two parts,
one in S and the other off S as follows
\begin{eqnarray}
C=\bar{C}+<C,N>N.
\end{eqnarray}
From the definition of $\textbf{N}$ and (7.1) we obtain
\begin{eqnarray}
N^{\alpha}=\frac{x^{\alpha}}{R}
\end{eqnarray}
Let us construct  the covariant derivative of $\textbf{C}$. We have
a local basis and  a diagonal and unitary tensor metric, so that the
Christoffel symbols are null. Then the covariant derivative of
$\textbf{C}$ in the $\textbf{Y}$ direction is given by
\begin{eqnarray}
\nabla_{Y}C=0.
\end{eqnarray}
It is easy to show that
\begin{eqnarray}
\nabla_{Y}N=\frac{Y}{R}.
\end{eqnarray}
The Lie derivative of the metric tensor in S is given by \cite{1},
\begin{eqnarray}
L_{\bar{U}}g'=2\lambda_{U}g',
\end{eqnarray}
where $\textbf{U}$ is a constant vector in the flat manifold, and
$\lambda_{U}$ is the  characteristic function. For S, the
characteristic function is given by
\begin{eqnarray}
\lambda_{U}=-\frac{1}{R}\{<U,N>.
\end{eqnarray}
Substituting (7.8) in (7.7) we have
\begin{eqnarray}
L_{\bar{U}}g'=-2\frac{1}{R}<U,N>g'.
\end{eqnarray}
In the region of S where $<U,N>$ is not null,  $ \bar{U} $ is a
conformal Killing vector and in the region where  $<U,N>$  is null,
 $\bar{U} $ is a Killing vector.
\newline
We now consider another constant vector $\textbf{V}$ in the flat
space. The Lie derivative of its projection in S is given by
\begin{eqnarray}
L_{\bar{U}}g'=-2\frac{1}{R}<U,N>g'.
\end{eqnarray}
As we consider a local basis and constant vectors $\textbf{U}$ and
$\textbf{V}$, the commutator is given by
\begin{eqnarray}
[U,V]=0.
\end{eqnarray}
Then,
\begin{eqnarray}
L_{[\bar{U},\bar{V}]}g'=-2\frac{1}{R}<[U,V],N>g'=0.
\end{eqnarray}
\newpage
Regardless  $ \bar{U}$ and $\bar{V}$ being Killing or conformal
Killing vectors, their commutator is a Killing vector. In the
following we will show that the commutator $[\bar{U},\bar{V}]$ is
proportional to the quantum angular momentum of a particle.
\newline
Using (7.3) in the following commutator of elements of the basis ,
we obtain
\begin{eqnarray}
\nonumber [\bar{U},\bar{V}]=\\
\nonumber =U^{\alpha}V^{\beta}[X_\alpha-<X_\alpha,N>N,X_\beta-<X_\beta,N>N]=\\
\nonumber =U^{\alpha}V^{\beta}[\bar{X}_\alpha,\bar{X}_\beta].\\
\end{eqnarray}
We now calculate the commutator of elements of the basis, by parts.
\newline
 We have by simple calculation
\begin{eqnarray}
<X_\alpha,N>N=\frac{1}{R}\eta_{\alpha\beta}x^{\beta}.
\end{eqnarray}
Substituting (7.14) in (7.13) we obtain
\begin{eqnarray}
\nonumber [\bar{X}_\alpha,\bar{X}_\beta]=\\
\nonumber =[X_\alpha,X_\beta]-[X_\alpha,\frac{1}{R}\eta_{\beta\sigma}x^{\sigma}N]+[X_\beta,\frac{1}{R}\eta_{\alpha\sigma}x^{\sigma}N]+\\
\nonumber +\frac{1}{R^2}[\eta_{\alpha\sigma}x^{\sigma}N, \eta_{\beta\sigma}x^{\sigma}N].\\
\end{eqnarray}
In a local basis we have
\begin{eqnarray}
[X_\alpha,X_\beta]=0,
\end{eqnarray}
\begin{eqnarray}
[\eta_{\alpha\sigma}x^{\sigma}N, \eta_{\beta\sigma}x^{\sigma}N]=0.
\end{eqnarray}
Substituting in (7.15) we obtain
\begin{eqnarray}
\nonumber [\bar{X}_\alpha,\bar{X}_\beta]=\\
\nonumber =
\frac{1}{R^2}(\eta_{\alpha\sigma}x^{\sigma}\frac{\partial}{\partial(x^{\beta})}-\eta_{\beta\sigma}x^{\sigma}\frac{\partial}{\partial(x^{\alpha})})\\
\nonumber =
\frac{1}{R^2}(x_\alpha\frac{\partial}{\partial(x^{\beta})}-x_\beta\frac{\partial}{\partial(x^{\alpha})})\\
\nonumber=-i\frac{1}{\hbar}\frac{1}{R^2}L_{\alpha\beta}.\\
\end{eqnarray}
Multiplying $ L_{\alpha\beta}$ by a vielbein basis  we obtain
\begin{eqnarray}
\nonumber
L_{(\mathbf{A})(\mathbf{B})}=\\
\nonumber
=(i\hbar)(R^{2})R_{(\mathbf{A})(\mathbf{B})(\mathbf{C})(\mathbf{D})}x^{(\mathbf{D})}\eta^{(\mathbf{C})(\mathbf{M})}\frac{\partial}{\partial(x^{\mathbf{M})}}.\\
\end{eqnarray}
where
\begin{eqnarray}
 \hat{p}_{(\mathbf{M})}=(i\hbar)\frac{\partial}{\partial(x^{\mathbf{M})}}
\end{eqnarray}
is the quantum momentum operator of a particle, and
\begin{eqnarray}
\nonumber
R_{(\mathbf{A})(\mathbf{B})(\mathbf{C})(\mathbf{D})}=\\
\nonumber
=\frac{1}{R^2}[\eta_{(\mathbf{A})(\mathbf{D})}\eta_{(\mathbf{B})(\mathbf{C})}-\eta_{(\mathbf{A})(\mathbf{C})}\eta_{(\mathbf{B})(\mathbf{D})}]\\
\end{eqnarray}
is the curvature of S in the vielbein basis  and $
\eta_{(\mathbf{A})(\mathbf{C})} $ is diagonal.
\newline
We consider as an important observation that the association between
the quantum angular momentum operator and the constant curvature
operator is allowed in an orthogonal vielbein basis of a Cartesian
coordinate, regardless of having a curved or a flat manifold. We
have used the embedding of a n-dimensional manifold S in an
n+1-dimensional flat manifold, only to obtain the quantum angular
momentum operator of a particle, without postulates.
\newline
We can rewrite (5.19) as follows
\begin{eqnarray}
\nonumber
L_{(\mathbf{A})(\mathbf{B})}=\\
\nonumber
=(i\hbar)[\eta_{(\mathbf{A})(\mathbf{D})}\eta_{(\mathbf{B})(\mathbf{C})}-\eta_{(\mathbf{A})(\mathbf{C})}\eta_{(\mathbf{B})(\mathbf{D})}].\\
\nonumber
.x_{(\mathbf{D})}\eta^{(\mathbf{C})(\mathbf{M})}\frac{\partial}{\partial(x^{\mathbf{M})}}.\\
\end{eqnarray}
\newpage
Note that the coordinates in (7.18) are in the n+1-dimensional
flat manifold  and  $L_{\alpha\beta}\subset S$, so that
$L_{\alpha\beta}=0$  for $\alpha$ or $\beta$ equal to $n+1.$
\newline
Racah has shown that \cite{10} the Casimir operators of  any
semisimple Lie group can be constructed from the quantum angular
momentum (5.22). Each multiplet of semisimple Lie group can be
uniquely characterized by the eigenvalues  of the Casimir operators.
\newline
Although we have built the quantum angular momentum from classical
geometric considerations we can write the usual expression for an
eigenstate of Casimir operator as follows
\begin{eqnarray}
\hat{C}\mid...>=C\mid...>.
\end{eqnarray}
In the following we calculate the Lie derivative of the so(p,n-p)
algebra. For the Lie group SO(p,q) we choose the signature
$(p,q)=(p,n-p)=(-,-,-,...-,+,+,..+)$, with the algebra
\begin{eqnarray}
\nonumber
[L_{(\mathbf{A})(\mathbf{B})},L_{(\mathbf{C})(\mathbf{D})}]=-i(\eta_{(\mathbf{A})(\mathbf{C})}L_{(\mathbf{B})(\mathbf{D})}+\eta_{(\mathbf{A})(\mathbf{D})}L_{(\mathbf{C})(\mathbf{B})}\\
\nonumber +\eta_{(\mathbf{B})(\mathbf{C})}L_{(\mathbf{D})(\mathbf{A})}+\eta_{(\mathbf{B})(\mathbf{D})}L_{(\mathbf{A})(\mathbf{C})}).\\
\end{eqnarray}
Considering the Lie derivative
\begin{eqnarray}
\nonumber \textbf{L}_{[L_{(\mathbf{A})(\mathbf{B})},L_{(\mathbf{C})(\mathbf{D})}]}g'=\\
\nonumber
=-\frac{1}{R}<[X_{(\mathbf{A})},X_{(\mathbf{B})}],[X_{(\mathbf{C})},X_{(\mathbf{D})}],N>g'=0,\\
\end{eqnarray}
where, for the orthogonal Cartesian coordinates, the vielbein is
given by
\begin{equation}
E_{\Lambda}^{(\mathbf{A})}=\delta_{\Lambda}^{(\mathbf{A})},
\end{equation}
we have
\begin{eqnarray}
\nonumber[X_{(\mathbf{A})},X_{(\mathbf{B})}]=[X_{\alpha},X_{\beta}]=0.\\
\end{eqnarray}
\newpage
Note that $g'$ in S is form-invariant  in relation to the
Killing's vector $\xi$ \cite{11} and  in relation to the  algebra of
SO(p,n-p)
 as well. We conclude that the algebra of SO(p,n-p) is a Killing's
object. The same is true for the algebra of the Lie group SO(n),
where for SO(n) we could  choose the signature  $(+,+,+...,+,+)$.
\newline
The constraint (7.1) is invariant for many of the classical groups.
For these groups it is possible to build operators, from the
combination of the quantum angular momentum operators, which are
Killing's objects in relation to $´g'$. Therefore, the metric is
form-invariant in relation to this algebra. It is interesting to see
some of these groups in the Cartan's list of irreducible Riemannian
globally symmetric spaces, \cite{5}, and in \cite{12}.
\newline
Note that we start from a normal coordinate transformation. In other
words, in the region where the transformation (2.4) is well-behaved,
we can build (3.10) and by a conformal transformation we have (5.23)
which was essential to obtain the quantum angular momentum operator
from geometry.


\newpage
\begin{thebibliography}{99}
\bibitem{1}W.C.Weber and S.I.Goldberg,{\it Queen's Papers in Pure and Applied Mathematics-No.16}
(Queen's University.Kingston.Ontaro,1969).
\bibitem{2} {\it in Lectures in Theoretical PhysicsXIII},
(A.O.Barut and W.E.Brittin, Eds.,Colorado Assoc. Universit, Boulder
Colo. 1971); R.Maartem and S.D.Maharaj,{\it Class.Quantum Grav.}
(3,1005,1986); N.V.Mitskievich and J.Horsky,{\it Class.Quantum
Grav.}(13,2603,1996);  A.J.Keane and R.K.Barrett, {\it Class.Quantum
Grav.}(17,201,2002); R.Banerjee,{\it Ann.Phys.} (311,245,2004).
\bibitem{3}E.Cartan,{\bf Le\c{c}\~ons sur la Geometrie des Espaces De Riemann},
(Gauthier-Villars, Paris,1946).
\bibitem{4}M.Spivak, {\bf A Comprehensive Introduction to Diferential Geometry}, {Volume two},
(Publish or Perish, Inc. 1999).
\bibitem{5}S.Helgason, {\bf Differential Geometry and Symmetric Spaces}
(Academic Press,1962)
\bibitem{6}O.Veblen and T.Y.Thomas, {\it Trans. Am. Math.Society.},(vol 25,551,1923).
\bibitem{7}L.P.Eisenart, {\bf Non Riemannian Geometry}
(Dover Publications,2005)
\bibitem{8}L.P.Eisenart, {\bf Riemannian Geometry}
(Princeton University Press,1997)
\bibitem{9}See reference [1]
\bibitem{10}W.Greiner and B.Muller, {\bf Quantum Mechanics Symmetries}
(Springer,1994).
\bibitem{11} S. Weinberg, {\bf Gravitation andCosmology:Principles and Applications of the General Theory of Relativity}
(John Wiley Sons1972)
\bibitem{12}R.Gilmore, {\bf Lie Groups,Lie Algebras, and Some of Their
Applications} (Dover Publications,Inc,2002)
\end{thebibliography}
\end{document}